\documentclass[%
 preprint,
 floatfix,
 amsmath,amssymb,
 aps,
]{revtex4-2}
\pdfoutput=1
\usepackage{graphicx}
\usepackage{dcolumn}
\usepackage[normalem]{xcolor}
\usepackage{soul}
\usepackage{bm}
\usepackage{amssymb}
\usepackage{amsthm}
\usepackage{amsmath}
\usepackage{miller}
\usepackage{textcomp}
\usepackage{gensymb}
\usepackage{siunitx}
\usepackage[utf8x]{inputenc}
\usepackage{multirow}
\usepackage{natbib}
\usepackage{ulem}

\newcommand*{\et}[0]{\textit{et~al.~}}
\newcommand*{\fig}[1]{Figure~\ref{fig:#1}}

\begin{document}

\title{The Origin of Photoplasticity in ZnS}

\author{Sevim Polat Genlik}
\altaffiliation{polatgenlik.1@osu.edu}
\author{Roberto C. Myers}
\author{Maryam Ghazisaeidi}
\email{ghazisaeidi.1@osu.edu}
\affiliation{%
  Department of Materials Science and Engineering, The Ohio State University, Columbus, Ohio 43210, USA.
}

\date{\today}

\begin{abstract}

ZnS is a brittle material but shows extraordinary plasticity during mechanical tests performed in complete darkness. This phenomenon is known as the photoplastic effect, whose underlying mechanisms have long been unclear. We study the impact of light, via photoexcited charge carriers, on the dislocation core structure and mobility using first-principles calculations. We calculate the core structure and the charge-dependent Peierls barriers of the glide set of Shockley partial dislocations in ZnS. Our findings reveal that locally charged dislocations capture excess carriers in the system, leading to core reconstructions that alter the Peierls barrier, resulting in higher barriers and lower mobility for these dislocations. This altered and asymmetric mobility, depending on dislocation character (edge or mixed) and local stoichiometry (Zn or S rich), is responsible for the brittle behavior of ZnS under light exposure and will be reversed in complete darkness.
\end{abstract}

\maketitle

\section{Introduction}

Photoplasticity is a brittle to ductile transition in materials under different light conditions. This behavior has been documented in several Group II-VI semiconductors, such as ZnS, which tends to break under white light but can endure strains up to approximately 45$\%$ in total darkness~\cite{oshima2018extraordinary}. 
Previous work on group II-VI and III-V inorganic compound semiconductors indicate the flow stresses of these materials can be affected by external stimuli such as light exposure\cite{koubaiti1997photoplastic,dong2022giant,oshima2018extraordinary}, electron injection\cite{maeda1984electron,maeda1996enhancement,faress1993tem} and applied electric field \cite{li2023harnessing, westbrook1962electromechanical,osip1986properties,faress1993tem}. 

Given that the mechanism for plasticity involves the creation and motion of dislocations within the crystal~\cite{taylor1934mechanism,polanyi1934gitterstorung,orowan1934plasticity}, these observations highlight that the mobility of dislocations can be influenced by external stimuli that change materials' carrier concentrations.

Controlling the mechanical properties of these materials with various external stimuli is promising, but a comprehensive understanding of the underlying mechanisms of these phenomena has remained elusive. Previously proposed models to interpret observations can be categorized into two. The first, recombination-enhanced dislocation glide (REDG)\cite{maeda1996enhancement}, suggests that non-radiative recombination of excited or injected electrons at dislocations creates localized lattice vibrations, assisting dislocation movement. The second model posits that charged dislocations are surrounded by a charge cloud of oppositely charged defects, which restricts their mobility\cite{whitworth1975charged}. While REDG model explains softening under light, it fails to explain why ZnS becomes brittle when exposed to light. Conversely, the charge cloud model explains ZnS hardening but not GaAs softening\cite{maeda1996enhancement,maeda1984electron}, despite the presence of charged dislocations in both materials. Therefore, investigating this phenomenon, particularly from a theoretical point of view, is crucial for understanding the principles behind the photoplastic effect.

Following the more recent report of the enhanced ductility of ZnS in complete darkness by Oshima \et ~\cite{oshima2018extraordinary}, this topic has gathered considerable renewed attention~\cite{li2023harnessing,oshima2020room,dong2022giant,shen2019modified,wang2019enhanced,ukita2019theoretical}, with several attempts to explain this behavior. While these studies provide significant steps towards understanding photoplasticity, the complete picture of the underlying mechanisms of this phenomenon has remained unclear. For example, two studies~\cite{shen2019modified,wang2019enhanced} attributed the observed transition from ductile to brittle behavior upon illumination to a change in the deformation mechanism, shifting from dislocation-dominated to twin-dominated, based on generalized stacking fault energy calculations. However, this explanation raises the more pivotal question: Why does the deformation mechanism alter under different lighting conditions? Matsunaga et. al. presented the first core structure calculations in ZnS using density functional theory (DFT)~\cite{matsunaga2020carrier}. They observed changes in the core structure of the dislocations and speculated that more stable structures should be more difficult to move, without direct calculations of the Peierls barrier.
While it is reasonable to assume that dislocations with different core geometry have different mobilities, core reconstruction energies alone are not sufficient surrogates for lattice resistance to dislocation motion.  Since the energetics of bond-switching and rearrangements during dislocation motion are the main factors affecting the dislocation mobility, a realistic explanation requires the direct calculation of charge-dependent Peierls barriers.

In addition, Li \et have studied the motion of dislocations in ZnS under electric field, which shares similarities with the photoplastic effect, in terms of the effect of non mechanical stimuli on dislocations in this material~\cite{li2023harnessing}. Experimental results show evidence for charged dislocations and an anisotropy in mobility of the dislocations. However, the computational work remains preliminary due to the fact that proper core reconstructions are not considered. As we show below, these reconstructions play a crucial role in determining the relative mobility of dislocations under different charge states.

Here, we discuss the impact of light, through photoexcited charge carriers, on the dislocation core structure and mobility, providing detailed mechanistic insight and quantitative analyses that were previously unexplored. We calculate the stable geometry and charge state for the glide set of Shockley partial dislocations in ZnS. Further analysis of the charge-dependent Peierls barriers indicates decreased mobility under light irradiation and a change in the degree of asymmetry in the mobility depending on the type (edge or mixed) and stoichiometry (Zn or S rich) of the partial dislocations. The rest of the paper is organized as follows. Section~\ref{sec:methodology},  details the computational methods. In Section~\ref{sec:results}, we present the calculated dislocation core structures and their charge-dependent properties, along with the analysis of the Peierls barriers. We also discuss these findings in detail and their implications for the photoplastic effect. Finally, in Section~\ref{sec:conclusions}, we conclude by summarizing the key outcomes of our study and highlighting the impact of light on the mechanical behavior of ZnS.

\section{Computational methods}

\label{sec:methodology}

All Density Functional Theory (DFT) calculations were performed with the Vienna ab initio Simulation Package (VASP) \cite{kresse1996efficient} using pseudopotentials generated by the projector augmented wave (PAW)  method \cite{kresse1999ultrasoft,blochl1994projector}. Exchange correlations were treated by Strongly Constrained Appropriately Normed (SCAN)\cite{sun2015strongly} Meta Generalized Gradient Approximation (meta-GGA) functional\cite{tao2003climbing}. A plane wave cut-off energy of 400 eV was used with a k-points density of 0.15 \r{A}$^{-1}$ for structural relaxations. The dislocations were introduced by displacing all atoms according to the elastic displacement field of the corresponding dislocation using anisotropic elasticity theory~\cite{anderson2017theory}. All atomic positions were subsequently optimized until forces are smaller than 10 meV/\r{A}$^{-1}$.

In this study, we employed full periodic boundary conditions with a quadrupolar arrangement of dislocations in a triclinic simulation cell with a dislocation dipole. To investigate core reconstructions, we used two sets of supercells with single and double lattice translation periods along the \hkl[1-10] dislocation line direction. The simulation cell consists of 576 atoms for double period and 288 atoms for single period dislocations. It is oriented along \hkl[-1-12], \hkl[111], and \hkl[1-10], corresponding to the x, y, and z directions, respectively. The $\{111\}$ planes of ZnS are polar because they consist of stacked layers of Zn and S atoms. Therefore, two possible types of edge-character dislocations can exist, in one of which the extra half-plane terminates with a row of Zn atoms, and in the other, with S atoms, resulting in a dislocation core region that is rich in Zn or S, respectively. For convenience in this paper, dislocations rich in S atoms at their core region will be referred to as `S-core,' and those rich in Zn atoms as `Zn-core'. As a result, our simulation cells with a dislocation dipole consist of pairs of one S-core and one Zn-core dislocation with a separation of approximately 24 \r{A} ~between them.

When ZnS is irradiated with light of a sufficient wavelength to induce inter-band transitions, photoexcited carriers become present within the crystal. We simulated the effect of photoexcited carriers by adding extra holes and electrons into the simulation cells and reoptimizing the atomic positions in their presence. In the 4-fold coordinated ZnS structure, each dangling bond requires 0.5 electrons to be saturated. Therefore, the charge added to the simulation cell was determined by  adding (or subtracting) 0.5 electrons per dangling bond within the dislocation core.

For the electronic band structure and density of states calculations only, we performed PBE+U calculations implemented in VASP using the Dudarev approach \cite{dudarev1998electron}. An effective Hubbard U correction was applied to both Zn-3d and S-3p states with values of 9 eV and 5 eV, respectively, ensuring the experimental band gap of ZnS, 3.8 eV, accurately reproduced.

\subsection{Stacking fault energy calculations}
Stacking fault energies were calculated using supercells composed of 30 $\{111\}$-type layers—15 Zn-type and 15 S-type—containing a total of 60 atoms to minimize interaction between the stacking fault and its periodic images. Stacking faults were introduced by tilting the $\langle111\rangle$ axis of the perfect zinc blende simulation cell along $[11\bar{2}]$ and $[\bar{1}10]$ directions on one of the narrowly spaced $\{111\}$ planes to obtain the  glide Generalized Stacking Fault Energy (GSFE) surface. The stacking fault energy is defined as:
\begin{equation}
\gamma = \frac{E_{\text{fault}} - E_{\text{perfect}}}{A}
\end{equation}
where \(E_{\text{fault}}\) and \(E_{\text{perfect}}\) represent the total energies of the system with and without a stacking fault, respectively, and \(A\) denotes the area of the glide plane over which the stacking fault is extended.

\subsection{Peierls barrier calculations}

We performed climbing image Nudged Elastic Band (cNEB) \cite{henkelman2000climbing,henkelman2000improved} calculations implemented in VASP to find the minimum energy path between two Peierls valleys as dislocations move between them. Three transition state images were used for interpolation between initial and final stable configurations. We utilized a triclinic supercell containing a dislocation dipole, as previously described, for these calculations. To ensure that the observed energy variation is exclusively attributable to changes in Peierls energy, maintaining a constant dislocation separation distance along the entire path is crucial\cite{pizzagalli2008calculations}. This approach guarantees that the elastic interaction energy remains unchanged. To maintain a fixed dislocation separation distance, the spring constant was set to 8 eV/\r{A}$^{2}$ during the relaxation until force convergence ($\leq$30 meV/\r{A}$^{-1}$) was achieved. Subsequently, to verify this,  the locations of dislocation centers were calculated using the following disregistry fitting after relaxing all c-NEB images:
\begin{equation}
\phi =\frac{b}{2\pi} \arctan\left(\frac{x-x_0}{\Delta}\right)
\end{equation}
where \( \phi \) is the disregistry, \( b \) is the Burgers vector, \( x_0 \) is the location of the dislocation center, and \( \Delta \) represents the core width.
 Then, separation distances were determined, and the maximum deviation is calculated as 5.1x10$^{-2}$ \r{A}, which was approximately ~2\% of the total displacement. This ensures the dislocation separation distance is kept fixed. 

\subsection{Core Energy Calculations}
The Energy Density Method (EDM) calculations were conducted using the EDM code\cite{yu2011energy}, implemented within the VASP frameworkto decompose total DFT energy into atomic energies. Energy and charge densities were computed on a real space grid of $672 \times 490 \times 96$. All EDM atomic energies were referenced against the bulk atomic energies of each species (Zn and S), calculated using the same method but in a perfect ZnS supercell without any dislocations. The elastic energy distribution around dislocations was analyzed by calculating the local atomic strain tensor using OVITO\cite{stukowski2009visualization}. These strains were obtained by comparing the deformed crystal structure with its perfect state. The elastic energy for each atom was determined using DFT-calculated stiffness constants (C$_{11}$=98.7GPa, C$_{12}$=62.2GPa and C$_{44}$=47.7GPa) with the formula:

\[
E_{\text{elastic}} = \frac{1}{2} \sum_{i,j,k,l} C_{ijkl} \varepsilon_{ij} \varepsilon_{kl}
\]

The total elastic energy within a cylindrical radius \(r\) was summed using:

\[
E_{\text{total}}^{\text{elastic}}(r) = \sum_{r_i \leq r} E_{\text{elastic}, i}
\]

Then the \(E_{\text{total}}^{\text{elastic}}(r)\) was plotted as a function of \(\ln(r)\), and deviations from linearity were identified to determine the core radius, marking the transition from the continuum elasticity region to the dislocation core region.

\section{Results and DISCUSSION}
\label{sec:results}

First, we calculate the core structure of relevant dislocations in ZnS. The predominant slip system in zinc blende crystals is  $\langle110\rangle\{111\}$. This crystal structure has two distinct types of $\{111\}$ planes: narrowly spaced and widely spaced.  Dislocations forming on narrowly and widely spaced $\{111\}$ planes are named glide and shuffle sets, respectively.  In covalently bonded semiconductors, glide sets of dislocations have lower formation energy~\cite{wang2006undissociated,cai2004dislocation} and are characterized as glissile~\cite{rabier2010chapter,masuda1983core}. Given the focus of this paper on plasticity, which is controlled by dislocation motion, we study the glide set of dislocations in ZnS.

\begin{figure}[t]
\includegraphics[width=8.6cm]{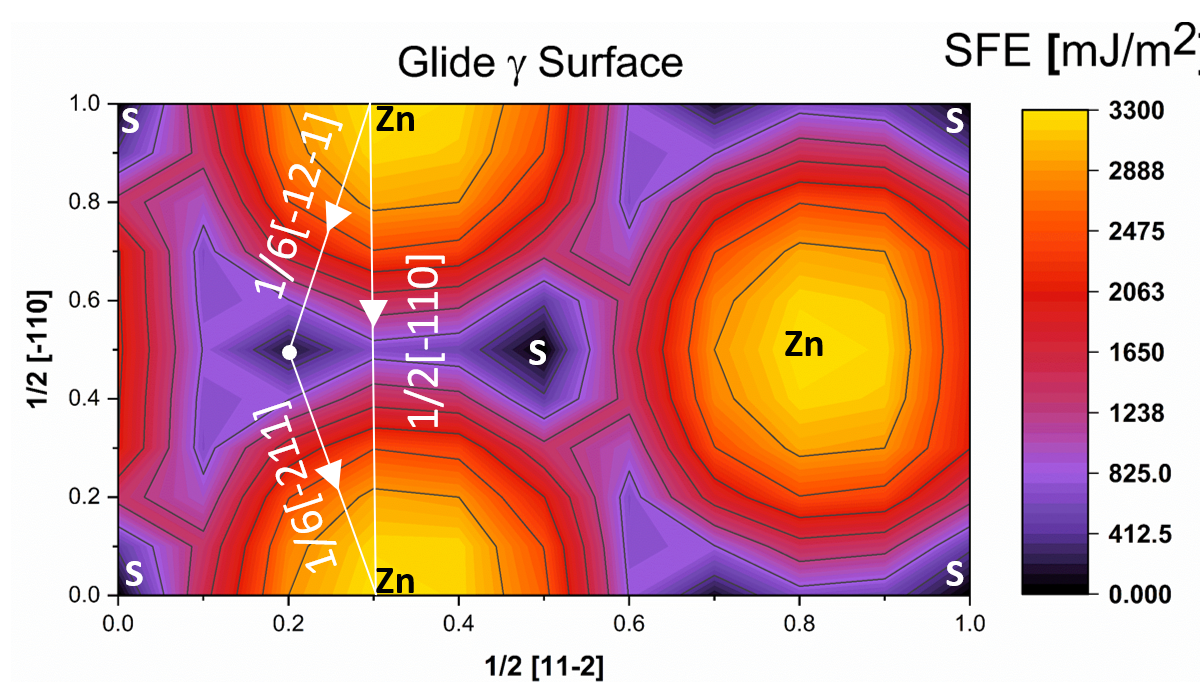}%
\caption{\label{fig:gamma_surface} The generalized stacking fault energy surface of a (111) plane in ZnS.  Arrows indicate the dissociation of $1/2[\bar{1}10]$ Burgers vector. The white dot shows an intrinsic stable stacking fault formed between two partials. 
}
\end{figure} 

There are some reports of the observation of partial dislocations in ZnS through transmission electron microscopy (TEM)\cite{faress1993mobility,faress1993tem,li2023harnessing}. Some also report the stacking fault energy (SFE) to be in the range of 6-10 mJ/m², as estimated by measuring the distance between observed partials in TEM images \cite{zaretskii1983isolated}. To theoretically confirm that dislocations dissociate in ZnS, we calculated the generalized stacking fault energy, or $\gamma$ surface, of the glide plane. The obtained $\gamma$ surface, as shown in Figure ~\ref{fig:gamma_surface}, indicates a stable intrinsic stacking fault with an energy of 12.1 mJ/m². This value is in good agreement with previous experimental estimates and proves that the SFE in ZnS is indeed low \cite{zaretskii1983isolated}, making dislocation dissociation feasible. Consequently, we proceeded to investigate the glide set of Shockley partial dislocations, specifically the mixed ($30^\circ$) and edge ($90^\circ$) glide Shockley partials.

These Shockley partials, with different numbers of dangling bonds, have non-stoichiometric cores, namely Zn-rich and S-rich, as discussed in the methods section. In Figure S2(a), we provide the excess electrostatic potential map for the 90° dislocations, indicating that the Zn core is positively charged and the S core is negatively charged. This can be explained by the electronegativity difference between Zn and S, where S has a greater attraction for electrons. \fig{edmmap90sp} (b) and (c) show the excess electrostatic potential maps with excess negative and positive charges, respectively. As expected, the excess negative charge is attracted to the positively charged Zn-core, and the excess positive charge to the negatively charged S-core, resulting in full neutralization.

For completeness, we include the core structures with a single periodicity along the dislocation line in the supplementary materials, emphasizing the importance of using a sufficiently large cell.  In the main part, we only focus on cores with double periodicity (DP) along the line direction as they are more realistic representations, allowing for accurate core reconstruction with lower core energies.

\begin{figure}
\includegraphics[width=\textwidth]{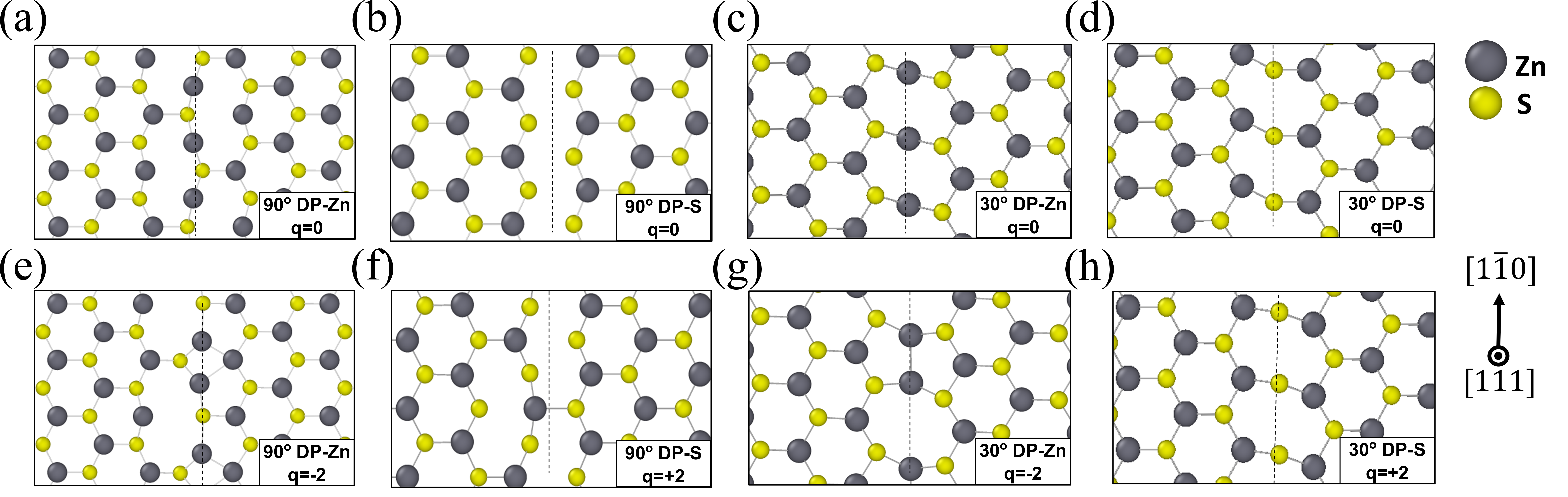}
\caption{\label{fig:relaxed_cores_DP} Core structures of the double period Shockley partial dislocations under different charge states. Atomic configurations for (a) 90° Zn-core and (b) 90° S-core, (c) 30° Zn-core, and (d) 30° S-core when there are no excess carriers in the simulation cell. Atomic configurations for (e) 90° Zn-core and (g) 30° Zn-core when there are two extra electrons and (f) 90° S-core and (h) 30° S-core when there are two extra holes in the simulation cell. The dislocation line is shown with a black dashed line.} 
\label{fig:wide}
\end{figure}

Figure ~\ref{fig:relaxed_cores_DP} demonstrates the impact of different charge states on the atomic arrangements within the dislocation cores. 
A consistent observation across all dislocations, regardless of their DP or SP classification (See Figure \ref{fig:relaxed_cores_sp} for SP configurations), is the presence of dangling bonds at their cores in the absence of excess carriers, suggesting instability of core structures. Unlike in pure elemental semiconductors with diamond cubic structure, such as diamond\cite{genlik2023dislocations,blumenau2002dislocations} or silicon \cite{kittler2011structure}, where period doubling results in the complete elimination of dangling bonds, DP dislocation cores in ZnS still possess unsatisfied bonds in their pristine state under a carrier-free environment (Figures ~\ref{fig:relaxed_cores_DP} (a-d)). This plays a critical role in the subsequent response of these dislocations to the introduction of excess carriers.

For 90$^{\circ}$ DP dislocations, the addition of extra electrons leads to reconstruction at the Zn core only (compare Figures ~\ref{fig:relaxed_cores_DP}(a) and (e)), while the addition of extra holes leads to reconstructions at only the S core (compare Figures ~\ref{fig:relaxed_cores_DP}(b) and (f)).

For 30° DP dislocations, the presence of two additional electrons again induces a reconstruction at the Zn core as evident from  Figures~\ref{fig:relaxed_cores_DP} (c) and (g). On the other hand, the S-core remained unchanged even with additional holes in the simulation, as seen in Figures~\ref{fig:relaxed_cores_DP} (d) and (h).

The S-core’s greater resistance to reconstruction is likely due to the distinct chemical natures of Zn and S. Zn favors metallic bonds in an electron-rich or neutral state when spatial distances are suitable. The flexibility of metallic bonds facilitates bond reconfiguration and reconstruction by allowing Zn atoms to adjust their positions more readily, leading to more pronounced structural rearrangement. In contrast, S is a non-metal with high electronegativity and a larger atomic radius. S-S bonding is characterized by covalent character, which is more directional and less flexible compared to metallic bonds, making bond reconfiguration more difficult. In addition, when S-rich dislocations capture holes, the resulting positive charge increases the electrostatic attraction repulsion between S atoms.
As a result, S-rich dislocations are less likely to undergo reconstruction upon capturing holes.

\begin{figure}
\includegraphics[width=\textwidth]{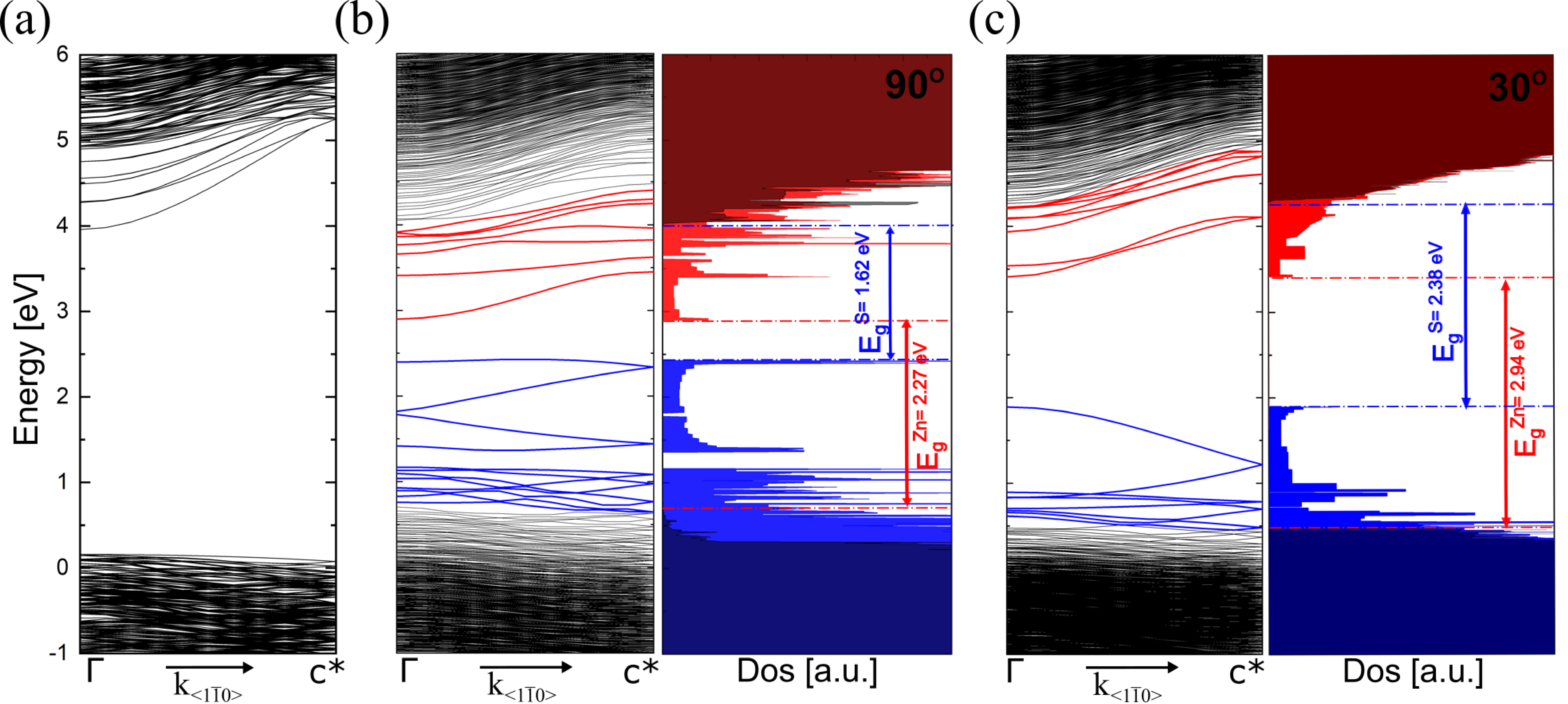}
\caption{\label{fig:bandgaps} Calculated electronic band structures along the dislocation line direction [1 -1 0] for (a) bulk reference, (b) 90$^\circ$ DP dislocation dipole, and (c) 30$^\circ$ DP dislocation dipole supercells in the absence of free carriers, accompanied by their respective total densities of states. The blue and red colors correspond to S and Zn cores, respectively. The black-shaded region represents the local density of states obtained from the atoms in the bulk-like region of the same supercells, serving as a reference}
\end{figure}

Figure~\ref{fig:bandgaps} shows the electronic band structures and density of states (DOS) plots of 90$^\circ$ and 30$^\circ$ DP partial dislocation dipole supercells, compared to the band structure of the bulk reference supercell in the same geometry as the dislocated ones. Defect-related states are color-coded according to the core stoichiometry, with Zn-core related defect states represented by red, the S-core related defect states represented by blue and the bulk-like (away from the dislocation cores) states represented by black. These electronic structure plots provide evidence that the states appearing within the forbidden band gap and close to the valence band edge originate from the S-core, while those near the conduction band edge originate from the Zn-core. The DOS plots of both types of dislocations exhibit the characteristic features of an ideal 1D Fermi gas~\cite{genlik2023dislocations}, as seen in Figure~\ref{fig:bandgaps} (b) and (c). Energy band gap values for each core, relative to the band edges of the local DOS of bulk-like regions, are provided in Table~\ref{tab:bandgap}.

\begin{table}[h]
\centering
\begin{tabular}{lcc}
\hline
\multicolumn{1}{c}{\textbf{Core Type}} & \multicolumn{2}{c}{\textbf{Partial Type}} \\ \cline{2-3}
\textbf{} & \textbf{90$^\circ$} & \textbf{30$^\circ$} \\ \hline
S & 1.62 & 2.38 \\
Zn & 2.27 & 2.94 \\ \hline
\end{tabular}
\caption{\label{tab:bandgap}Energy band gap values (in eV) for 90$^\circ$ and 30$^\circ$ Double Period S-Core and Zn-Core dislocations, in reference to valence band and conduction band edges obtained from the local density of states from the bulk-like region of dislocation dipole supercells under carrier-free conditions.}
\end{table}

The bulk ZnS band gap of 3.8 eV is too large for absorption of light in the visible part of the spectrum. However, as seen from Table~\ref{tab:bandgap}, dislocations in ZnS reduce the band gap so that it overlaps with the energy range of visible light (approximately 1.6 to 3.1 eV). The defect states within the band gap are spatially localized to the dislocation core, but delocalized (dispersing) along the dislocation line direction. The reduced bandgap within the dislocation cores now spans the visible light region, enabling direct photocarrier generation within the dislocations. The relevant transitions are shown in Figure 3. In the case of Zn core dislocations, the optical transitions, primarily involving blue-green light absorption, generate a photoelectron confined to the dislocation core, and a free valence band hole in the ZnS, whereas for S core dislocations, optical absorption, involving red-green light absorption, generates a photo-hole confined to the core, and a free electron in the conduction band of ZnS (See Figure \ref{fig:cartoon} (a)). If these calculations are accurate, then ZnS, which is transparent, should develop color as the density of dislocations increases due to plastic deformation. More specifically, three of the four dislocations are predicted to have an absorption onset between 422 and 546 nm (blue to green visible light), leading to a predicted orange-red hue under white light illumination. Interestingly, this anticipated coloration is observed in Figure 1 of Oshima \et \cite{oshima2018extraordinary}, but no discussion or explanation of that was offered.Results from the same paper indicate that at higher strain values, ZnS samples exhibit a more reddish color, suggesting increased plasticity likely due to the Zn cores (primarily involving blue-green light absorption) contributing to ductility more prominently. This suggests that Zn cores not only move more readily in darkness but also have a significant effect on plasticity, which will be later supported by our mobility calculations. When the light is turned off, the photo-carriers will recombine, returning the dislocations  to their original charged state. This process is likely to cause blue-green photoluminescence (PL) emission (See Figure \ref{fig:cartoon} (b)). Retrieving the original charge state of the dislocatfion can also explain why the photoelastic efect is reversible, as observed in experiments.

\begin{figure}
\includegraphics[width=\textwidth]{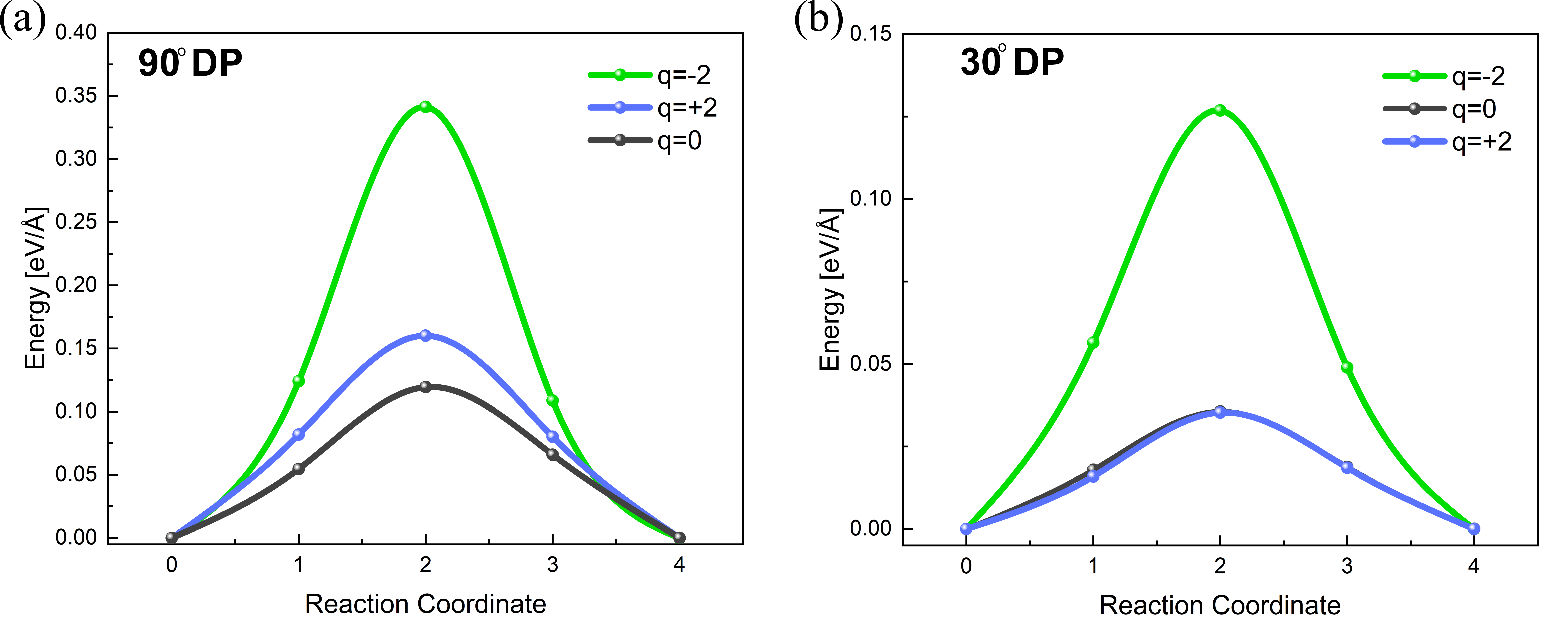}
\caption{\label{fig:Peierls} Peierls Barrier of a) 90° DP and b) 30° DP glide Shockley partials under different charge states. Energy variation (in eV/A) is shown as a function of reaction coordinate between two Peierls valleys.
} 
\label{fig:wide}
\end{figure}

Next, we study the effect of the various observed core reconstructions on the mobility of the dislocations by calculating the Peierls barrier, which
represents the energy barrier for a straight dislocation  to move between adjacent valleys in the potential energy landscape of the lattice \cite{anderson2017theory}.
\fig{Peierls} shows the Peierls barriers corresponding to the motion of the dislocation dipoles under various charge conditions, calculated by the nudged-elastic-band method.
The results show that the core reconstructions induced by carrier capture at the dislocation cores increase the Peierls barrier, thereby decreasing the dislocations' mobility.

In addition, the comparison of Peierls  barriers of the 90$^\circ$ and 30 $^\circ$ DP dislocations indicates that the 90$^\circ$ dislocations  exhibit lower mobility under both carrier-free or carrier-rich conditions. Also, as stated before, the structural analysis of the core geometry reveals that the S-core of 30$^\circ$ dislocation does not undergo reconstruction upon the addition of holes. Consequently, the almost identical core structure results in the same Peierls potential for neutral and hole-rich supercells, as shown in Figure \ref{fig:Peierls} (b).

Another significant conclusion to be drawn is that the addition of extra electrons, resulting in Zn-core reconstructions, causes a substantial increase in the Peierls barrier compared to S-core reconstructions (Figure \ref{fig:Peierls} (a)). This indicates that Zn-core dislocations have a more pronounced impact on ZnS's photoplastic behavior.  At the same time, it should be noted that due to the use of periodic boundary conditions, we considered pairs of 90$^{\circ}$ S-core and 90$^{\circ}$ Zn-core (or 30$^{\circ}$ S-core and 30$^{\circ}$ Zn-core) dislocation dipoles, within a single simulation cell. Consequently, the calculated Peierls barriers correspond to the energy required to move two dislocations. Although a single type of carrier induces reconstruction in only one core (S or Zn) while leaving the other unaltered, it is not straightforward to allocate the obtained energy barrier between them, given the different geometry of the cores.

\begin{figure}[h]
\includegraphics[width=\textwidth]{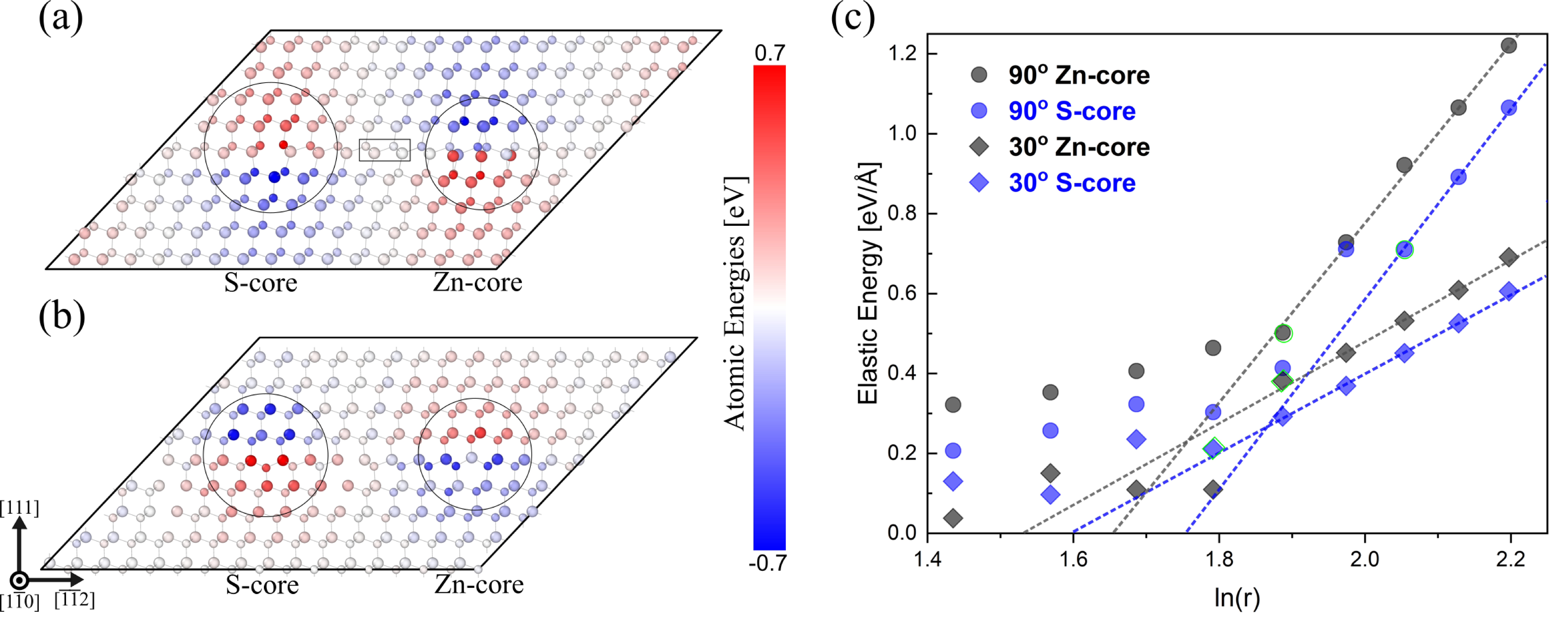}
\caption{\label{fig:edmatomic} Atomic energy distribution maps for  (a) 90$^\circ$ and (b) 30$^\circ$ DP dislocation cores. Atoms are color-coded based on their EDM-calculated energies relative to the bulk atomic energies for each species computed using the same method but in a perfect supercell without dislocations. The dislocation line runs perpendicular to the page. The stacking fault region, utilized for calculating the EDM-derived stacking fault energy, is marked with a black box at the center of the dislocations. (c) The total elastic strain energy, within a cylindrical volume, plotted against the logarithm of that cylinder radius (r). The core radii, identified at points deviating from the linear zone, are highlighted with green boxes and circles on the graph. Core radii determined for each core are also depicted in supercells (a) and (b) as black circles.} 
\end{figure}

Next, we use the energy density method (EDM) \cite{dan2022first,yu2011energy} to partition the total DFT-computed energies into atomic energies, thereby computing each dislocation's core energy and Peierls barrier separately. The atomic energy distribution maps of the 90$^\circ$ and 30$^\circ$  DP dislocation dipole supercells are shown in Figures \ref{fig:edmatomic}(a) and (b), respectively. Atoms with energies significantly higher or lower than the bulk atomic energies form two distinct clusters, each localized to the core of the two dislocations forming the dipole.

The total dislocation energy per unit length,  within a cylinder of radius $r$, is the sum of the elastic energy and the core energy, $E_{\text{core}}$ defined as
\begin{equation}
\label{eq:lineEnergy}
E_{\text{disloc}} (r) = \frac{Kb^2}{4\pi} \ln\left(\frac{r}{r_0}\right) + E_{\text{core}},
\end{equation}
where $r_0$ indicates the core's estimated radius, $b$ is the Burgers vector, and $K$ is the energy prefactor determined by the elastic constants of the material \cite{hull2011introduction}. 

To estimate the core radii of each dislocation, the elastic energy was first calculated within a cylindrical volume of radius \(r\). These energies were then graphically represented as a function of \(\ln(r)\), as shown in Figure \ref{fig:edmatomic}(c) for 90$^\circ$ and 30$^\circ$ DP dislocations. Since the elastic energy is expected to be linear with \(\ln(r)\) in the material's elastic region away from the core, a zone exhibiting this linear behavior was identified. Least squares fitting was applied to the data within this linear zone. Then, the core radius was estimated based on the deviation of the actual data from the fitted linear trend. Finally, the sum of the atomic energies within the cylindrical volume of the estimated core radius was computed for each dislocation core and reported as the core energy in table \ref{tab:coreprops}
\begin{table}[h]
\centering
\begin{tabular}{lccc}
\hline
\textbf{Partial Type} & \textbf{Core Type} & \textbf{Core Radius [\r{A}]} & \textbf{Core Energy [eV/\r{A}]} \\
\hline
\multirow{2}{*}{90$^\circ$} & Zn & 6.59 & 0.089 \\
                            & S  & 7.77 & 0.098 \\ \hline
\multirow{2}{*}{30$^\circ$} & Zn & 6.00 & 0.074 \\
                            & S  & 6.60 & 0.060 \\ \hline
\end{tabular}
\caption{\label{tab:coreprops}Core radii and EDM calculated core energies for 90$^\circ$ and 30$^\circ$ dislocations' Zn and S cores under carrier-free conditions.}
\end{table}

Additionally, for verification purposes, the total EDM energy of atoms in the stacking fault region (depicted by the black box in Figure \ref{fig:edmatomic} (a)) is calculated and then scaled to a unit area to determine the stacking fault energy. This value is found to be 9.8 mJ/$m^{2}$, which aligns closely with the directly calculated value of 12.1 mJ/$m^{2}$, indicating a reasonable agreement between the two methods.

\begin{table}[ht]
\centering
\begin{tabular}{lcccccc}
\hline
\textbf{Partial Type} & \multicolumn{3}{c}{\textbf{90$^\circ$ [eV/\AA]}} & \multicolumn{3}{c}{\textbf{30$^\circ$  [eV/\AA]}} \\
\cline{2-4} \cline{5-7}
 & \( q=0 \) & \( q=-2 \) & \( q=+2 \) & \( q=0 \) & \( q=-2 \) & \( q=+2 \) \\
\hline
\( [Zn] \) & 0.034 & 0.275 & 0.029 & 0.013 & 0.099 & 0.013 \\
\( [S] \) & 0.102 & 0.097 & 0.115 & 0.027 &0.034  &  0.021\\
\( [Zn]+[S] \) & 0.136 & 0.372 & 0.144 &  0.040& 0.133 & 0.034 \\
\( [Zn+S] \) (NEB) & 0.119 & 0.341 & 0.160 & 0.036  & 0.127 & 0.035 \\
\hline
\end{tabular}
\caption{\label{tab:edmpeierls}Calculated core-resolved Peierls barrier values (in eV/\r{A}) for 90$^\circ$ and 30$^\circ$ DP dislocations under various supercell carrier concentrations using Energy Density Method (EDM). The last row shows the combined Peierls barrier magnitude obtained directly from Nudged Elastic Band (NEB) calculations for comparison. The 'q' values indicate the quantity and polarity of excess carriers in the system.}
\end{table}

Next, we use the same methodology to determine the  Peierls barriers corresponding to the individual 90$^\circ$ and 30$^\circ$  dislocations. This is done by calculating the core energy difference between the stable configuration and the saddle point, i.e., the minimum and maximum points on the minimum energy paths provided in Figure \ref{fig:Peierls}, respectively. 

Table \ref{tab:edmpeierls} shows the charge-dependent Peierls barrier corresponding to Zn- and S-rich partial dislocations, revealing  a significant change upon carrier trapping at the dislocations. Most notably, the reconstruction of the 90$^\circ$  Zn core following the capture of two electrons leads to an increase in the Peierls Barrier from 0.034 eV/Å to 0.275 eV/Å. In contrast, the S-core reconstruction triggered by the capture of two holes results in only a modest increase of the barrier from 0.102 eV/Å to 0.115 eV/Å. These findings indicate that the impact of light illumination—or any other external stimuli that alter the carrier concentration within the material—on dislocation mobility is asymmetric for different types of dislocations. Particularly, the Zn core dislocations, for both mixed and edge types, exhibit a substantial reduction in mobility, with the Peierls Barrier escalating to approximately eight times its original value subsequent to carrier capture. Note that the Peierls barrier values calculated using EDM are overestimated compared to those calculated using NEB in general. This discrepancy can be attributed to the core radius being merely an estimate, and slight changes to the core size change the corresponding core energy.

Overall, the core resolved Peierls barriers indicate: (1) Pure edge (90$^\circ$) dislocations are less mobile than 30$^\circ$ dislocations having mixed character irrespective of the carrier concentrations in the system. (2) The presence of excess carriers in the system reduces the mobility of 90$^\circ$ and 30$^\circ$  Zn-core dislocations, as well as 90$^\circ$ S-core dislocations, but it has no effect on 30$^\circ$ S-cores. (3) The most pronounced increase in the Peierls Barrier is observed in Zn-Core dislocations, irrespective of whether they have a pure edge or mixed character. These three conclusions collectively indicate that there is an asymmetry in the mobility of partial dislocations, depending  on their character (edge or mixed) and stoichiometry (Zn-rich or S-rich).

Moreover,  two  previous studies\cite{shen2019modified,wang2019enhanced}, have reported that the deformation mechanism of ZnS shifts from slip-dominated to twin-dominated upon irradiation with light, based on the analysis of calculated stacking fault energy curves.  It is also previously reported that the prominent low-temperature deformation mode in ZnS and most of the compound semiconductors with sphelarite structure is mechanical twinning \cite{vanderschaeve1998mechanical}. Deformation twins in face-centered cubic (fcc) crystals are formed by  the glide of Shockley partial dislocations on successive \hkl{111} planes. Consequently, the twinning system in fcc materials is of the \hkl{111}\hkl<112> type. 
Asymmetry in the mobility of leading and trailing partials can cause an uneven advancement of partial dislocations in the slip plane and provide favorable conditions for twin nucleation. Our results suggest that twin deformation is likely to occur under both dark and light irradiation conditions due to the asymmetry in dislocation mobilities observed in both scenarios. However, under light, there exists a greater degree of asymmetry in dislocation mobilities, which is likely to promote twinning compared to slip in ZnS.

\section{Conclusion}
\label{sec:conclusions}
We investigated the underlying mechanisms behind the photoplastic effect observed in ZnS.
First, we studied the effect of photo-excited carriers on the core structure of the glide set of Shockley partial dislocations.  We showed that dislocation Zn- and S-rich cores have positive and negative charges, respectively, which locally alter the band structure and lead to the trapping of the oppositely charged carriers into the dislocation cores. Upon attracting the photo-excited charge carriers, the dislocation cores undergo structural changes. 
Subsequently,  calculation of the charge-dependent Peierls barrier revealed that: (1) Dislocations with pure edge character exhibit lower mobility compared to mixed type dislocations. (2) The presence of excess carriers reduces the mobility of  both 90° and 30° Zn-rich dislocations, and the 90° S-core dislocation. However, it has no effect on the mobility of the 30° S-rich dislocations. (3) The most pronounced effect on mechanical behavior changes, through alteration in dislocation mobility, originates from dislocations that are rich in Zn atoms at their core region, regardless of their type (edge or mixed). (4) There exists an asymmetry in the mobility of partial dislocations and the degree of this asymmetry increases under light. (5) The increase in both 90° and 30° dislocation mobility and the decreased asymmetry in the mobility of the 90° vs. the 30° partial dislocations are responsible for the ductile behavior of ZnS in darkness.
These findings  offer mechanistic insight into the  photoplastic effect observed in ZnS and its deformation modes under varying lighting conditions. 

\begin{acknowledgments}
We gratefully acknowledge the support of this work by the AFOSR grant number FA9550-21-1-0278. Computational resources were provided by the Ohio Supercomputer center. We thank Dallas Trinkle for sharing their Energy Density Method code.
\end{acknowledgments}

\bibliographystyle{apsrev4-2}
\bibliography{bibfile2}

\newpage
\section{ Supporting Information}
\renewcommand{\thefigure}{S\arabic{figure}}
\setcounter{figure}{0}
\subsection{ Single Periodicity (SP) Dislocation Core structures}

\begin{figure} [h]
\includegraphics[width=\textwidth]{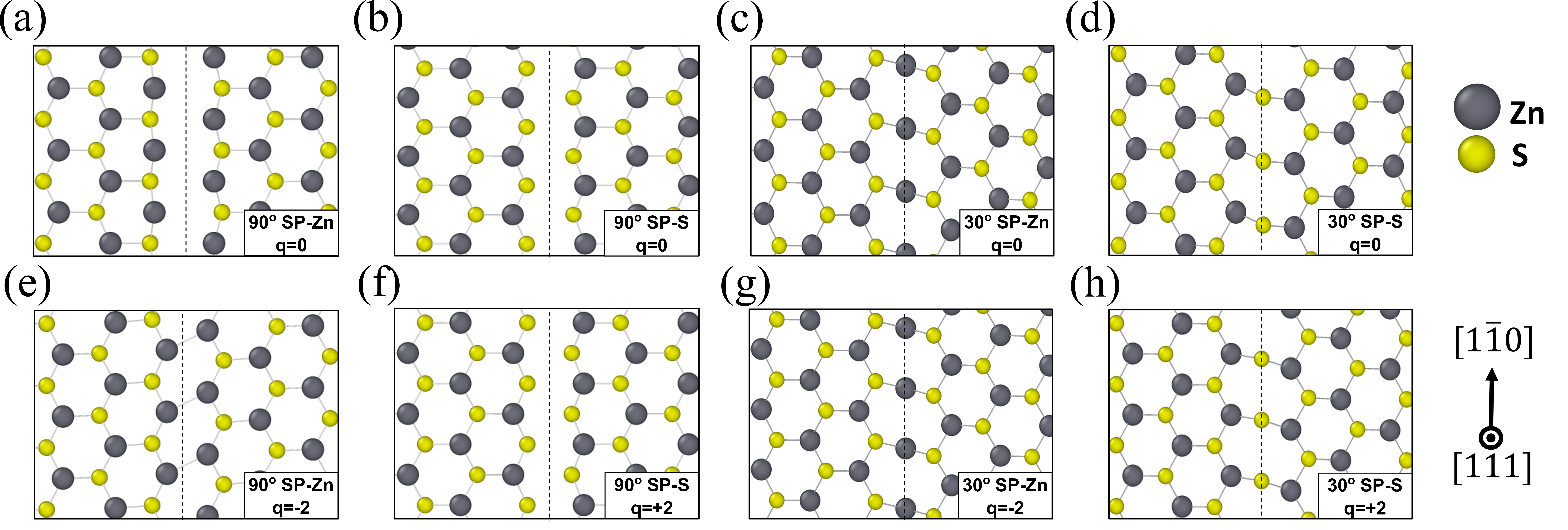}
\caption{\label{fig:relaxed_cores_sp} Cross-sectional views of relaxed single period (SP) dislocation core structures under varying carrier conditions in the simulation cell. a) 90° Zn-core, b) 90° S-core, c) 30° Zn-core, and d) 30° S-core show the atomic configurations without excess carriers. e) 90° Zn-core and g) 30° Zn-core display configurations with two additional electrons in the simulation cell, while panels f) 90° S-core and h) 30° S-core show configurations with two extra holes.The dislocation line is shown with a black dashed line.} 
\end{figure}
For SP dislocations, the 30$^{\circ}$ type exhibits no reconstructions within either core upon the introduction of excess carriers (Figure \ref{fig:relaxed_cores_sp}(c,g) and (d,h)). The 90$^{\circ}$ type only shows reconstructions at the Zn core when extra electrons are present in the system (Figure \ref{fig:relaxed_cores_sp}(a,e)). This reconstruction occurs through the rotation of Zn atoms with dangling bonds at the core along the [1-10] axis, causing a decrease in the Zn-Zn distance on either side of the dislocation center from approximately 3.95 \r{A} to 2.71 \r{A}. Therefore, Zn-Zn distances become comparable to the experimentally observed Zn-Zn metallic bond length of 2.66 \r{A} \cite{wyckoff1963interscience}. This leads to the formation of metallic bonds between them.

\subsection{Impact of Excess Carriers on Dislocation Local Charges, Bond Lengths, and Orientations"}

\begin{figure}[h]
\includegraphics[width=\textwidth]{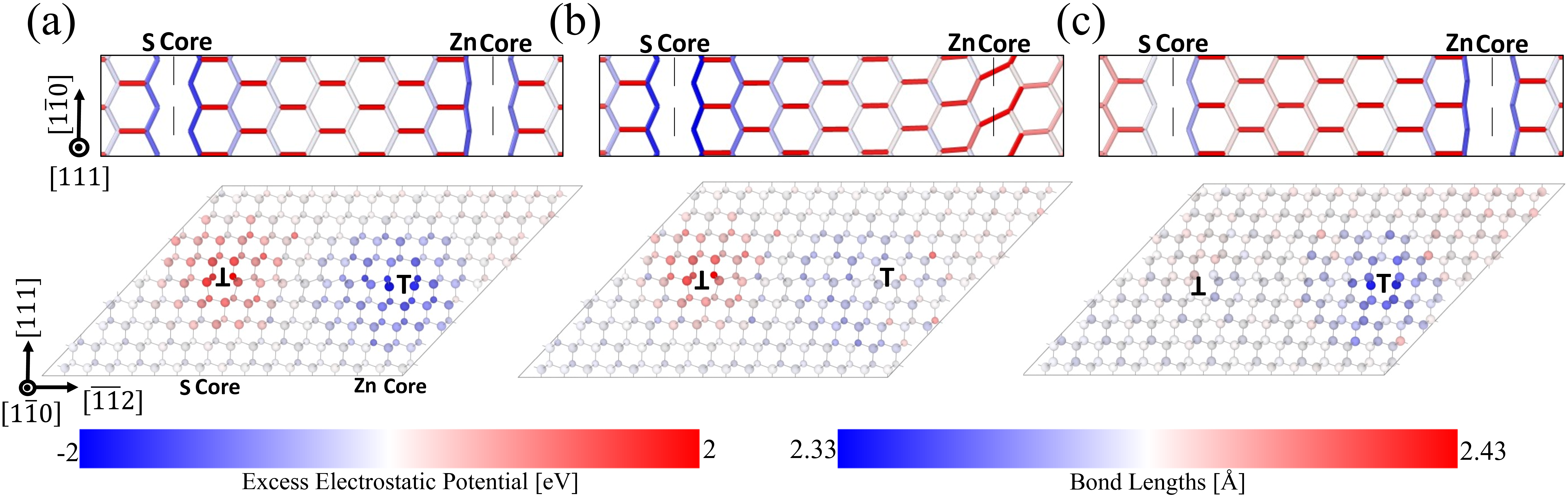}
\caption{\label{fig:edmmap90sp} Dislocation cores under different charge states. Top Panels (a-c) present the cross-sectional view of the relaxed core structures of a single period (SP) 90$^{\circ}$ dislocations under various charge states, with atomic bonds color-coded based on their length deviations from the bulk bond length of ZnS (2.37 Å). Panel (a) represents the neutral state, (b) shows the structure with excess negative carriers, and (c) illustrates the structure with excess positive carriers. Bottom Panels (a-c): Excess electrostatic potential maps of 90$^{\circ}$ SP dislocation dipoles under various charge states. Atoms are color-coded according to excess electrostatic potentials. Bottom panel (a) depicts the neutral state with charged dislocations, (b) displays the structure with negative excess carriers, and (c) shows the structure with positive excess carriers} 
\end{figure}

\begin{figure}[h]
\includegraphics[width=\textwidth]{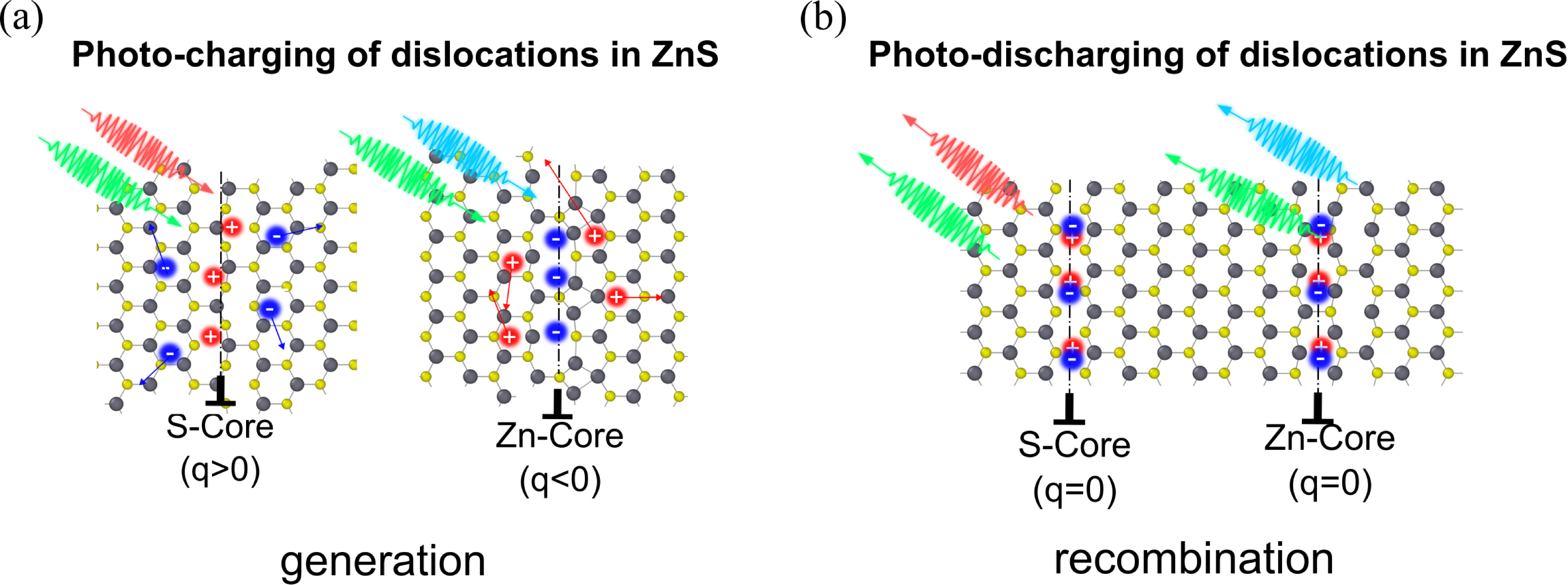}
\caption{\label{fig:cartoon} Schematic representation of optical absorption and carrier dynamics in ZnS dislocations. (a) For Zn core dislocations, blue-green light absorption generates a photoelectron confined to the dislocation core and a free valence band hole in the ZnS bulk. For S core dislocations, red-green light absorption generates a photo-hole confined to the dislocation core and a free conduction band electron in the ZnS. With confined carriers at the core, the dislocations become neutral. (b) Upon turning off the light, the recombination process leads to the dislocations being charged again as the photoexcited carriers recombine} 
\end{figure}

In scenarios where the obvious reconstructions are not observed, such as in 30$^{\circ}$ DP, 30$^{\circ}$ and 90$^{\circ}$ SP S-cores in the presence of extra holes in the system, significant alterations in bond orientations and lengths were observed. For example, for 90$^{\circ}$ SP case, we analyzed both the excess electrostatic potential maps and bond length distributions before and after carrier additions. The potential maps reveal that the S-core, initially locally negative in a neutral system, captures extra holes, thereby neutralizing the core region ( Figure \ref{fig:edmmap90sp} (a),(c),bottom). 
Similarly, the Zn-core, being locally positive in a neutral system, captures electrons as shown in Figure \ref{fig:edmmap90sp}(a),(b),bottom. Upon analyzing the deviation of Zn-S bond lengths at the core region from their bulk value, we observed that the initially compressed bond lengths within the S-core approach the bulk value of $~$2.37 \r{A} upon capturing extra holes ( Figure \ref{fig:edmmap90sp} (a),(c),top). This suggests a relaxation of the bonds towards the bulk value upon neutralization, indicating the release of elastic energy previously stored in the compressed bonds.

\subsection{Light-Induced Carrier Dynamics in Dislocation Cores}

\end{document}